\documentclass[11pt,a4paper,twoside]{article}
\usepackage[utf8]{inputenc}
\usepackage{amsmath}
\usepackage{amsfonts}
\usepackage{amssymb}
\usepackage{graphicx}
\usepackage{authblk}

\author{Sebastian Jaksch \footnote{Corresponding Author: S. Jaksch, s.jaksch@fz-juelich.de}}
\author{Alexandros Koutsioubas}
\author{Stefan Mattauch}
\author{Olaf Holderer}
\author{Henrich Frielinghaus}
\affil{Forschungszentrum J\"ulich GmbH, JCNS at Heinz Maier-Leibnitz Zentrum, Lichtenberstraße 1, 85747 Garching, Germany}

\title{Preliminary Report on Measurements of Dynamic Contributions to Coherent Neutron Scattering}

\begin{document}
\maketitle

\begin{abstract}
During this experiment we were testing the hypothesis that standing waves in a phospholipid membrane stack indeed lead to a detectable signal in coherent grazing-incidence small-angle neutron scattering, GISANS. These modes were identified earlier in a previous experiment using grazing-incidence neutron spin-echo spectroscopy, GINSES, (Jaksch, S., Frielinghaus, H. et al. (2017). Nanoscale rheology at solid-complex fluid interfaces. Scientific Reports, 7(1), 4417.). In order to identify those modes and prove conclusively that they were indeed a dynamic mode of the membrane and not a measurement artifact we were following a predetermined protocol: Starting at a physiological temperature (35$^\circ$C), where the modes were previously identified in GINSES, we lowered the temperature of the sample. Dynamic modes as an eigenmode of the lamellar system would under those conditions at least shift the position of any occurring peaks. Possibly below a phase transition temperature (approximately 25$^\circ$C for such a layered system) the peaks due to coherent scattering from a standing wave would vanish altogether, as the standing wave cannot be sustained by thermal excitation at that temperature. Upon reheating, any scattering contribution that is purely due to scattering from a standing wave would reappear. All experiments were performed on the MARIA instrument at MLZ.
\end{abstract}
\newpage

\section{Sample}
The sample was L-$\alpha$-phosphatidylcholine (SoyPC) on polished silicon blocks. Before deposition the Si-blocks were cleaned via the RCA procedure, which also resulted in a hydrophilic surface of the blocks. In our previous manuscripts we used an identical preparation process. \cite{Ibu,GINSES}  The SoyPC was dissolved in isopropanol p.A. and shaken for 15 minutes, until a clear solution with a slight yellow tint was obtained. Afterwards, the block was kept in a scaffold, that kept liquid from running off its surface and the obtained solution was cast on top of the silicon block. In order to gently dry the SoyPC film the block with the SoyPC solution on top was kept in a vacuum oven at room temperature and a pressure of 25\,kPa, just over vapour pressure of the solution. The drying procedure was performed over night and after 12 h the vacuum pressure was reduced to below 5\,kPa in order also to remove residual solvent.

Immediately after drying the block was transferred to a sample holder, screwed shut and the sample holder was filled with D$_2$O. Afterwards the sample holder was transferred to MARIA, connected to a Julabo thermostat and the experiment was started.

\section{Protocol and Instrument Setup}
As previously described we wanted to investigate the standing waves that were identified by GINSES \cite{GINSES} in an earlier experiment by means of GISANS. As we were aware that these features would necessarily be very close to the primary beam at low $Q$-values, we decided to use a pencil-shape primary beam with the best available resolution of 1$\times$1\,mm at MARIA. We were aware that under those conditions probably Bragg scattering from the membranes would be negligible, however we expected the features due to the standing wave to appear primarily around the specularly reflected beam in in-plane direction. All measurements were performed under an incident angle of 0.2$^\circ$ and lasted for 6\,h per temperature.

As we wanted to prove the dynamic character of the standing wave mode, we adapted a temperature protocol that would most likely lead to a change in position or intensity of features connected to dynamic scattering. Therefore we started at a temperature where we already knew that the modes were present from the GINSES measurement (37$^\circ$C). After that we gradually decreased temperature to 25 and 20$^\circ$C and the reheated to 37$^\circ$C. As there is a phase transition in the bulk phase of SoyPC at 25$^\circ$C \cite{phase}  we anticipated a drastic change or vanishing of the scattering contribution by the standing wave below that temperature and their reappearance upon reheating.

\section{Overview of the results}
\begin{figure}
\centering
\begin{minipage}[t]{0.45\textwidth}
\large{a)}

\includegraphics[width=\textwidth]{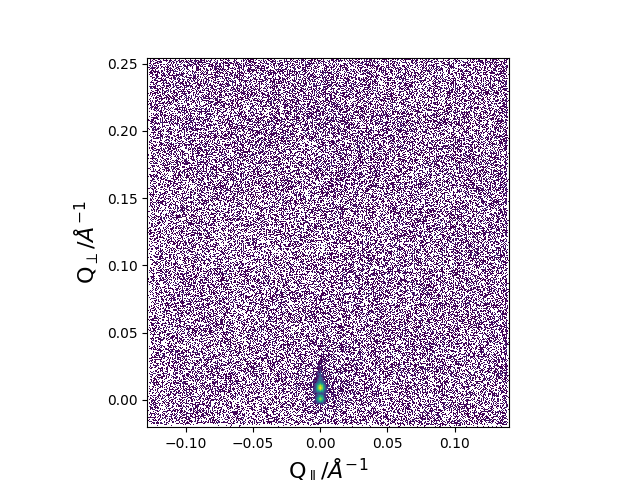}
\end{minipage}
\begin{minipage}[t]{0.45\textwidth}
\large{b)}

\includegraphics[width=\textwidth]{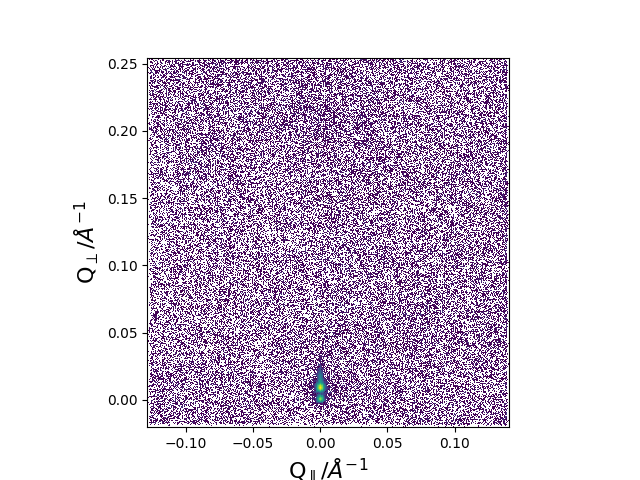}
\end{minipage}

\begin{minipage}[t]{0.45\textwidth}
\large{c)}

\includegraphics[width=\textwidth]{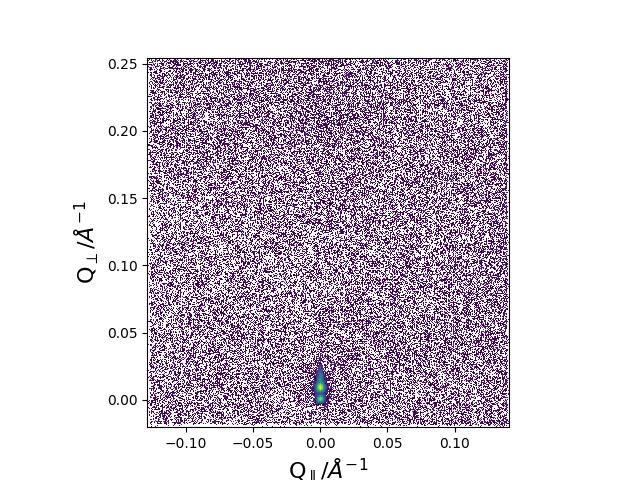}
\end{minipage}
\begin{minipage}[t]{0.45\textwidth}
\large{d)}

\includegraphics[width=\textwidth]{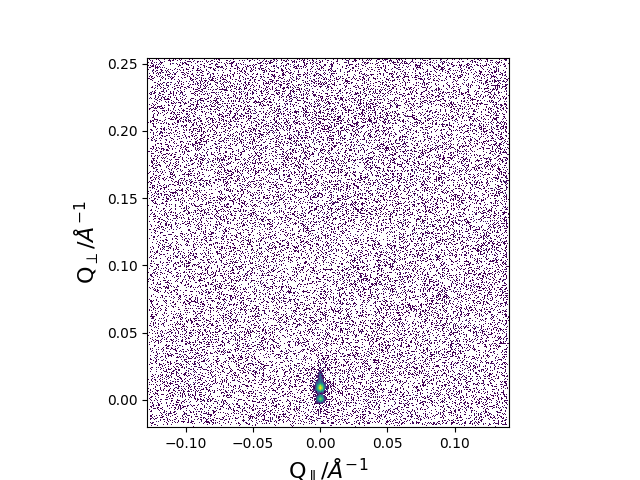}
\end{minipage}
\caption{GISANS images for a) 37$^\circ$C initial conditions, b) 25$^\circ$C, c) 20$^\circ$C, d) 37$^\circ$C final conditions.}
\label{fig:GISANS-images}
\end{figure}

The GISANS images are shown in Fig.\ref{fig:GISANS-images}. As anticipated there is no Bragg scattering due to the lamellar nature of the sample as was reported in earlier experiments. This feature should appear above the central primary and specularly reflected beam at approximately $Q_\perp = 0.1$\AA$^{-1}$.

With the bare eye the different GISANS images are near indistinguishable, but anyway the anticipated signal would be very small.

\begin{figure}
\centering
\includegraphics[width=0.7\textwidth]{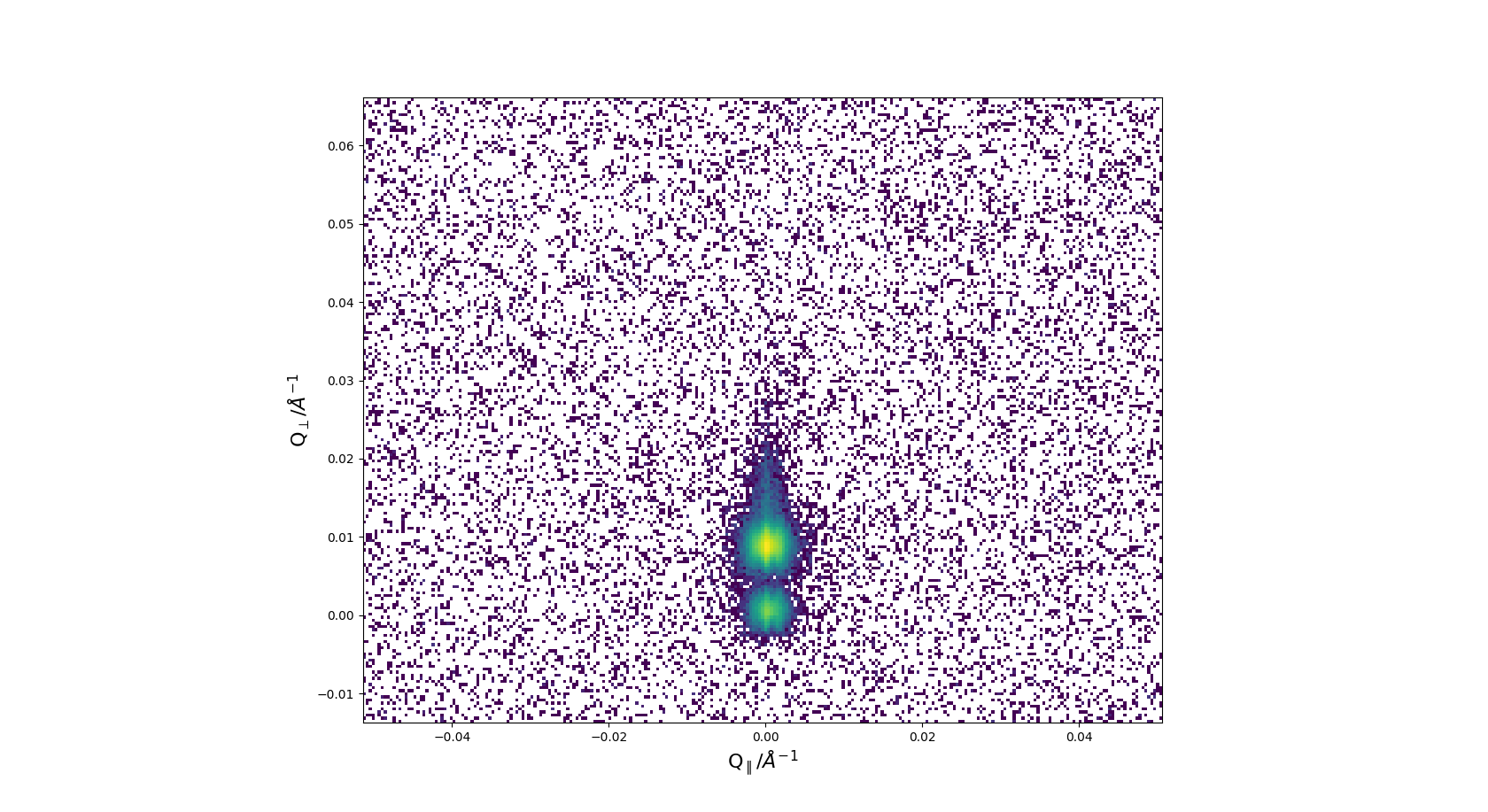}

\caption{GISANS images for 37$^\circ$C final conditions, zoomed in into the region around primary and specularly reflected beam.}
\label{fig:GISANS-ZoomIn}
\end{figure}

A zoom-in in Fig.\ref{fig:GISANS-ZoomIn} reveals there is a slight tail to the top of the specularly reflected beam, which may be due to out-of-plane decaying correlations, which are not further considered in this report. In the region of interest identified by the GINSES experiment of $Q_\parallel\approx0.01$\AA$^{-1}$, there is however no parasitical or other feature that might impede data analyis.

\section{Preliminary Data analysis}
In order to identify weak scattering contributions at $Q_\parallel\approx0.01$\AA$^{-1}$ we performed in-plane cuts through the specularly reflected peak. In order to improve statistics we averages over both directions from the primary beam as well as over a height of 8 pixels. In the in-plane direction the average was taken over 5 data points. We made sure the features discussed here were not an averaging artifact, however the average chosen here shows the discussed features in the clearest manner we could obtain by varying the average boundaries.

\begin{figure}
\centering
\includegraphics[width=0.7\textwidth]{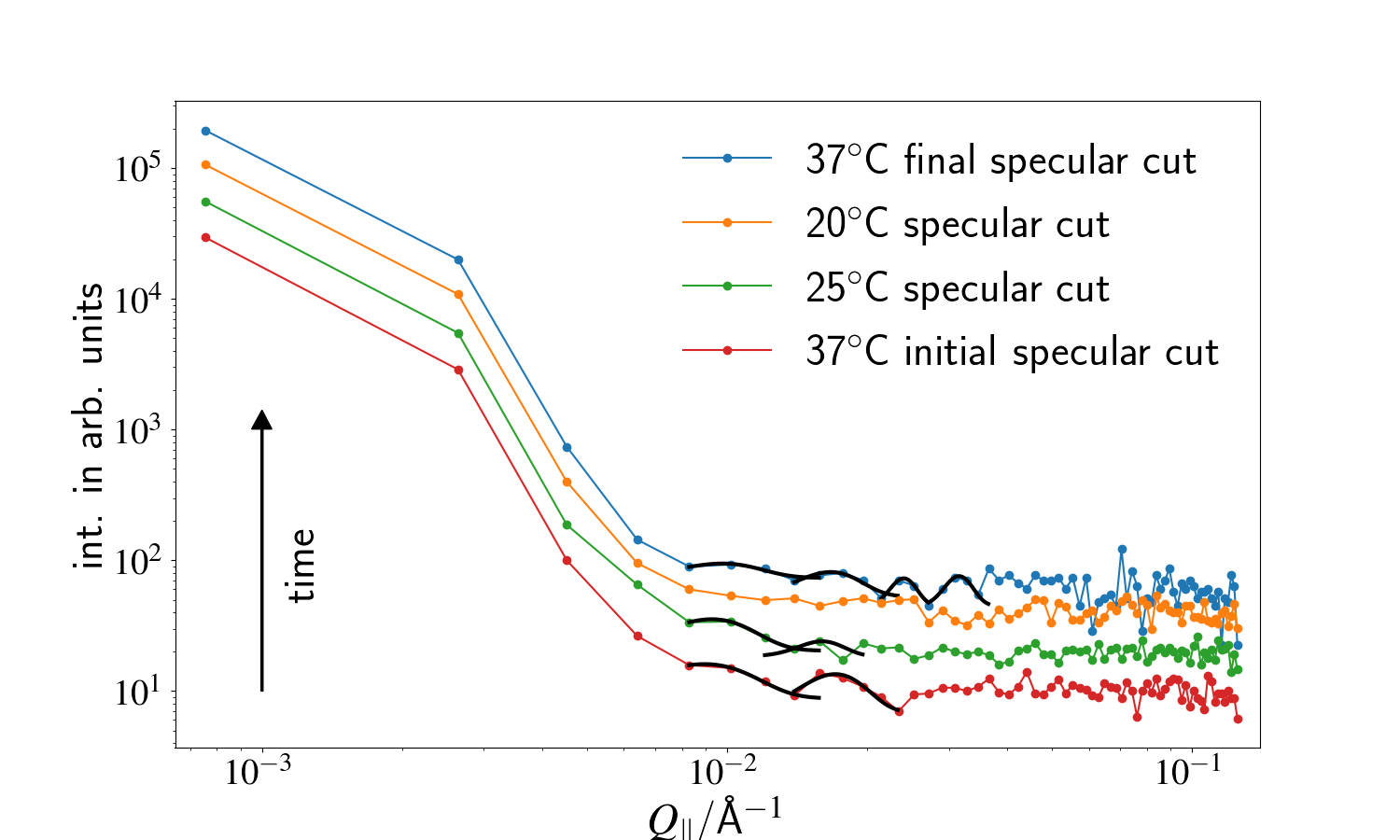}

\caption{Specular cuts in the in-plane direction of the different GISANS images shown above. Black lines show Gaussian fits to identify peaks around the region of interest. Intensities are shifted for better visibility by factors of 2 (25$^\circ$C), 4 (20$^\circ$C), and 16 (37$^\circ$C, final).}
\label{fig:specularCuts}
\end{figure}

The resulting cuts are shown in Fig.\ref{fig:specularCuts}. Peaks as expected from scattering from a standing wave in the surface of the membrane occur in the region of interest. From our previous GINSES experiment we expected a feature at $Q_\parallel\approx0.009$\AA$^{-1}$ at 37$^\circ$C. However, due to the geometry and the fact that $Q_\perp$ and $Q\parallel$ are intrinsically linked in GINSES, where me measure only one $Q$ and then have to determine the in-plane component by hand this value cannot be considered as exact. Therefore, the peak at $Q_\parallel\approx0.01$\AA$^{-1}$ can be exactly identified with the expected scattering from a standing wave. Additionally to the expected peak at $Q_\parallel\approx0.01$\AA$^{-1}$ also one higher order peak is visible.

The same is true for the sample at 25$^\circ$C. Also here two peaks can be identified, and they are indeed closer together that at higher temperatures ($\Delta Q_\parallel\approx0.007$\AA$^{-1}$). This is reasonable, since one can assume that at lower temperatures thermal excitations will carry less energy, hence have a longer wavelength. This directly translates into a smaller distance in $Q$.

At 20$^\circ$C degrees the previously identified dynamic peaks indeed vanish and therefore proof that either the excitation is also vanished or at least has moved out of the sensitivity window of the method.

Also as anticipated, the peaks reappear when the sample temperature is returned to 37$^\circ$C. The spacing is again identical to the one at the initial conditions ($\Delta Q_\parallel\approx0.01$\AA$^{-1}$), albeit now three higher order peaks become visible. It can also reasonably be assumed that longer training (here 24 h) improves the overall ordering of the system and therefore improves also the conditions needed for a collective in plane mode to form such a standing wave.

\section{Preliminary Discussion and Outlook}
Regarding the goal at the beginning of the experiment we were able to fully confirm our hypothesis, that standing waves identified by GINSES in phospholipid membrane stacks can indeed also be identified by GISANS. The occuring dynamic peaks behave as expected for scattering from thermally excited eigenmodes/standing waves. In a different lamellar phase they disappear and the physical range/wavelength of the standing waves can be determined by setting different temperatures.

Using this knowledge we have a powerful tool to combine the time consuming GINSES measurements with coherent GISANS experiments. While a GINSES experiment typically takes up to 5 days for a single $Q$-value, a GISANS experiment can be performed in the same amount of hours, resulting in a time gain of approximately one order of magnitude. This will allow us to scan larger areas of the phase space (temperature and sample composition), identify interesting areas in the phase space and then subsequently perform more accurate measurements using GINSES. This is a huge advantage, since for most of the samples, where GINSES is an option, anyway it is sensible to perform GISANS measurements in advance. However, to obtain data as shown here it is essential that the collimation settings  are tuned in a manner that allow the analysis of the specular reflection afterwards.

From the material science approach to this specific biologic sample we can speculate that, as the observed dynamic peaks and therefore also the standing waves, disappear or change drastically outside a physiological temperature window that these modes contribute to the biological function of the cell membrane. This can include trans-membrane transport mechanism such as cytosis, which is of paramount importance both in terms of pharmaceutical/biological application development as well as fundamental understanding of cellular mechanisms.   

\bibliography{literatureReport}{}
\bibliographystyle{ieeetr}

\end{document}